\documentclass[runningheads]{llncs}
 \usepackage{mathrsfs}
 \usepackage{amssymb}
 \usepackage{multirow}
\usepackage{hyperref}
\usepackage{graphicx}
\usepackage{enumitem}
\usepackage{marvosym}
\usepackage{amsmath}
\usepackage[table, svgnames]{xcolor}

\begin{document}
\title{Correlation-Distance Graph Learning for Treatment Response Prediction from rs-fMRI\thanks{
This research is supported in part by the Beijing Hospitals Authority Youth Program (ref: QML20191901), Beijing Hospitals Authority Clinical Medicine Development of Special Funding (ref: ZYLX202129), Beijing Hospitals Authority’s Ascent Plan (ref: DFL20191901), Training Plan for High Level Public Health Technical Talents Construction Project (ref: TTL-02-40), Research Cultivation Program of Beijing Municipal Hospital (ref: PZ2023032), EPSRC NortHFutures project (ref: EP/X031012/1).}}

\titlerunning{Correlation-Distance Graph Learning for Treatment Response Prediction}
%

\author{Xiatan Zhang \inst{3}\orcidID{0000-0003-0228-6359} \and
Sisi Zheng \inst{1,2} \orcidID{0000-0002-1575-6877} \and 
Hubert P. H. Shum \inst{3}\textsuperscript{(\Letter)} \orcidID{0000-0001-5651-6039} \and
Haozheng Zhang \inst{3}\orcidID{0000-0003-1312-4566} \and
Nan Song \inst{1,2} \and
Mingkang Song \inst{1,2} \orcidID{0000-0001-8146-1543}  \and \\ 
Hongxiao Jia \inst{1,2}\textsuperscript{(\Letter)} \orcidID{0000-0002-4924-1388}}
\authorrunning{Zhang et al.}
%
\institute{
Beijing Key Laboratory of Mental Disorders, National Clinical Research Center for Mental Disorders $\&$ National Center for Mental Disorders, Beijing Anding Hospital, Capital Medical University, Beijing, China\\
\and
Advanced Innovation Center for Human Brain Protection, Beijing Anding Hospital, Capital Medical University, Beijing, China  \\
\email{zhengsisi@ccmu.edu.cn} \\
\email{sss11nn@163.com} \\
\email{songmed003@mail.ccmu.edu.cn}\\
\email{jhxlj@ccmu.edu.cn} 
\and
Department of Computer Science, Durham University, 
Durham, United Kingdom\\
\email{\{xiatian.zhang, hubert.shum, haozheng.zhang\}@durham.ac.uk}
}

\maketitle  
\begin{abstract}
Resting-state fMRI (rs-fMRI) functional connectivity (FC) analysis provides valuable insights into the relationships between different brain regions and their potential implications for neurological or psychiatric disorders. However, specific design efforts to predict treatment response from rs-fMRI remain limited due to difficulties in understanding the current brain state and the underlying mechanisms driving the observed patterns, which limited the clinical application of rs-fMRI. To overcome that, we propose a graph learning framework that captures comprehensive features by integrating both correlation and distance-based similarity measures under a contrastive loss. This approach results in a more expressive framework that captures brain dynamic features at different scales and enables more accurate prediction of treatment response. Our experiments on the chronic pain and depersonalization disorder datasets demonstrate that our proposed method outperforms current methods in different scenarios. To the best of our knowledge, we are the first to explore the integration of distance-based and correlation-based neural similarity into graph learning for treatment response prediction.
\end{abstract}

\section{Introduction}
Resting-state functional magnetic resonance imaging (rs-fMRI) detects spontaneous fluctuations in the blood-oxygen-level dependent (BOLD) signal of the brain \cite{Khosla2019}. It provides insights into functional connectivity (FC) between different brain regions, aiding the understanding of neurological disorders \cite{Du2018}. However, the dynamic nature of neural patterns may restrict their use in identifying treatment biomarkers and predicting responses \cite{Taylor2021}. Leveraging deep learning to infer treatment outcomes from rs-fMRI is a potential solution \cite{Kong2021}, and establishing a framework that considers the current brain states and their underlying mechanisms could increase the clinical relevance of rs-fMRI.

Traditional FC analysis uses Pearson correlation to represent neural similarity, a key component in understanding brain region interactions \cite{Khosla2019}. However, recent studies suggest that both correlation and distance-based similarity metrics offer distinct advantages in representing neural similarity across different tasks and brain states \cite{Bobadilla-Suarez2020}. In particular, distance-based measures have been shown to enhance machine learning performance in rs-fMRI analysis \cite{Xiao2022}. That reveals the potential value of integrating distance-based similarity measures for deciphering complex neural mechanisms underlying observed patterns, which may also potentially improve the treatment response prediction.

FC naturally exhibits a graph structure \cite{Wang2010}, making graph representation and graph neural networks (GNNs) effective for capturing complex relationships in brain networks \cite{Zhou2020}. Recent studies have introduced graph representation and GNNs into FC analysis, grounded on two perspectives of FC \cite{Yu2018}: static FC graph (SG) \cite{Kan2022}, which assumes constant FC within a scan, and dynamic FC graph (DG) \cite{Dahan2021,Gadgil2020,Kim2021}, which assumes varying FC within a scan. With the non-stationary nature of rs-fMRI FC \cite{Chang2010}, DG approaches may be potentially more suitable for FC analysis in rs-fMRI. A recent exemplar is the Spatio-Temporal Attention Graph Isomorphism Network (STAGIN) proposed by Kim et al. \cite{Kim2021}, which applies a Graph Isomorphism Network (GIN) \cite{Xu2018} with an attention mechanism to model the spatio-temporal interplay between ROIs and their divided time windows. However, these studies have primarily concentrated on diagnosis rather than predicting treatment responses, and their graph representations exclusively depend on correlation-based similarity measures.

Several studies have utilized machine learning techniques to predict treatment responses from rs-fMRI FC data. Cao et al. \cite{Cao2020} employed support vector machines (SVMs) \cite{Cortes1995} to predict the treatment response of schizophrenia. Similarly, Kong et al. \cite{Kong2021} utilized spatial-temporal graph convolutions to predict treatment response for major depressive disorder, outperforming conventional machine learning methods. However, these approaches are still solely based on the correlation-based FC for rs-fMRI and cannot infer the distinctive temporal signal segment associated with the treatment response, thereby providing an incomplete understanding and interpretability of the complex spatiotemporal FC dynamics underlying treatment response in rs-fMRI.

In this work, we aim to handle the challenge of predicting treatment response in rs-fMRI which needs comprehending more intricate underlying mechanisms compared to other status-understanding tasks.  We present a framework utilizing both correlation-based and distance-based similarities to capture complex brain dynamics, beyond conventional Pearson correlation-based FC methods. We introduce a dynamic Correlation-Distance Graph Isomorphism Network (CD-GIN) with contrastive loss, enhancing spatio-temporal feature learning across varied time points. Further, we incorporate a Convolutional Block Attention Module (CBAM) \cite{Woo2018} to highlight critical graph representations, aiding precise inference. Our method, evaluated on a chronic knee osteoarthritis (OS) clinical trial dataset \cite{Tetreault2016} and a real-world depersonalization disorder (DPD) treatment dataset, demonstrates the robustness and generalizability of our proposed framework. Source code is available on \textcolor{blue}{\url{https://github.com/summerwings/CDGIN}}.

The contributions of this work are summarized as follows:
\begin{enumerate}
\item To the best of our knowledge, we are the first to explore the integration of distance-based and correlation-based neural similarity into graph learning for treatment response prediction, finding a complementary relationship between the two streams from our experiments on two datasets.
\item We present the Correlation-Distance Graph Isomorphism Network (CD-GIN) with a contrastive loss function to dynamically learn unique spatio-temporal features from different similarity measures, improving the modeling of complex brain dynamics.
\item We integrate the Convolutional Block Attention Module (CBAM) to identify and interpret key graphs and time windows in rs-fMRI, enhancing predictive accuracy and uncovering significant predictors of treatment response.
\item We provide a new open benchmark for predicting treatment response in chronic OS, using clinical trial data, and construct a real-world dataset for DPD treatment, demonstrating our approach's generalizability.
\end{enumerate}

\section{Related Work}
\subsection{Distance-based Neural Similarity}
Several studies have leveraged distance-based neural similarity in machine learning for rs-fMRI analysis. Xiao et al. introduced distance measures to capture voxel-wise time course spatial relations within ROIs, leading to distance-based covariance descriptors used for correlation computation and age prediction \cite{Xiao2022}. Ma et al. extended this work using distance-based and Pearson correlations to predict age \cite{Ma2023}.  Despite these advancements, such studies primarily used correlation analysis to establish FC, without fully utilizing distance measures. This may overlook regional signal dissimilarities and non-linear ROI relationships. They also considered only static FC for single rs-fMRI scans, possibly neglecting FC temporal dynamics. These intricacies, vital for predicting treatment responses \cite{Del2023}, inspire our work to directly employ distance measures to dynamically represent FC, aiming to capture complex spatio-temporal ROI relationships.

\subsection{Dynamic Graph Learning for fMRI}
Several studies have explored dynamic graph representation for FC in machine-learning-based fMRI analysis. For example, Gadgil et al. \cite{Gadgil2020} utilized Spatio-Temporal Graph Convolutional Networks (STGCN) \cite{Yan2018} to extract spatial-temporal features, and Kim et al. \cite{Kim2021} introduced STAGIN, combining a GIN \cite{Xu2018} with an attention mechanism. 
Although these studies were primarily focused on sex classification, their graph representations align with the dynamic nature of neural activity \cite{Chang2010}, a property that is also crucial for treatment response prediction. However, these methods solely relied on correlation-based similarity measures and could miss key distance-based and non-linear ROI information. In contrast, our work integrates distance-based FC into dynamic learning, uniquely positioned to capture both linear and non-linear brain dynamics, enhancing its potential for treatment response prediction.

\section{Methodology}
\begin{figure*}[h]
\centering
\includegraphics[scale = 0.185]{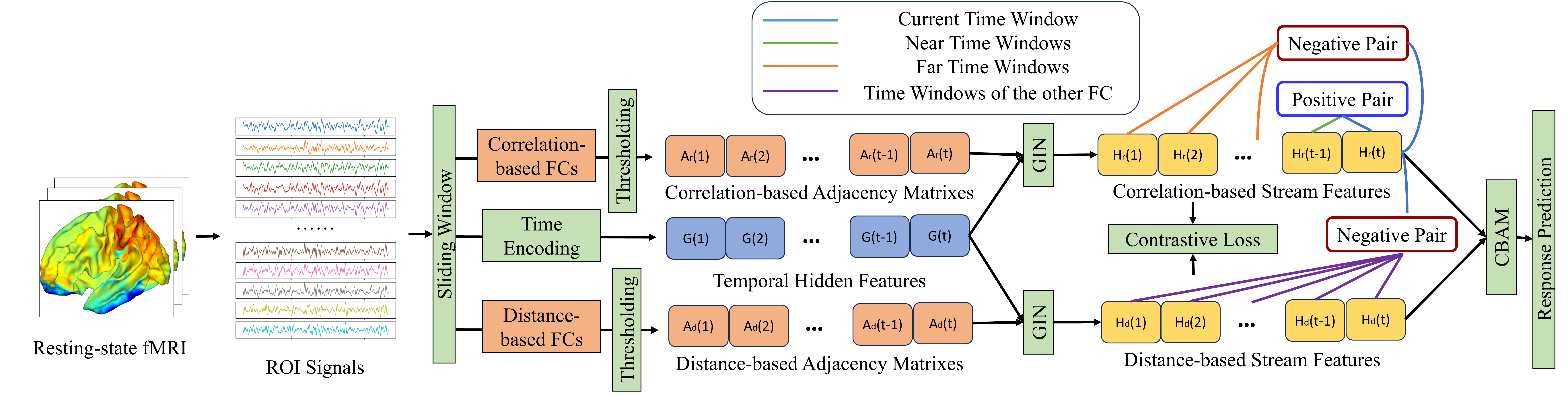}
\caption{The Framework of Correlation-Distance Graph Learning: The displayed contrastive pairing pertains to one example time window. However, in practical implementation, the contrastive loss is actually paired and calculated for each time window.}
\label{fig:Net}
\end{figure*}

In our proposed framework (Fig. \ref{fig:Net}), signals from predefined ROIs in rs-fMRI are extracted to capture regional activity. The dynamic FC graph is then constructed using both correlation and distance-based similarity measures, representing complex brain dynamics. The CD-GIN, incorporating a contrastive loss, models spatio-temporal features and learns distinctive information between different FC and time windows. Finally, the CBAM is employed to weigh key graphs and time windows to get the probability for prediction, which refines inference and enhances interpretability.

\subsection{Dynamic Functional Connectivity Graph based on Correlation and Distance-based Similarity}

\noindent \textbf{Task Formulation}
We employ a pre-defined 3D atlas \cite{Faria2012} to partition the brain into distinct ROIs, extracting ROI signal sequences 
$X \in \mathbb{R}^{T\times M}$ from the input sequence. $T$ is the fMRI data length and $M$ is the atlas-specified ROI number. We formulate treatment response prediction as a classification task, aiming to predict response $y$, a binary variable determined by the main clinical efficacy. It assesses the reduction rate of relevant symptom rating scores to determine whether the treatment is effective for the corresponding diseases \cite{Tetreault2016}. Our proposed network infers the treatment response $\hat{y}$ from the ROI signal sequences $X(t)$ over different time windows $t$, where $t$ denotes a sub-sequence extracted from $X$ to capture evolving patterns in rs-fMRI \cite{Chang2010}.

\noindent \textbf{Dual-Stream Neural Similarity for Dynamic Functional Connectivity}
We propose an integration of distance-based and traditional correlation-based measures for fMRI FC analysis, aiming to capture both linear and non-linear interactions among brain regions \cite{Janse2021}. The integration of distance-based measures arises from the limitations of correlation-based FC, which often overlooks essential non-linear interactions that influence treatment efficacy. This oversight is partially attributed to a bias induced by stimuli causing large activations \cite{Walther2016}. Distance-based similarities, by their nature, assume that entities closer within a metric space share higher similarity, making them suitable for revealing these overlooked non-linear relationships. To align with correlation-based measures, where higher values typically indicate greater similarity, we represent similarity using negative distances. Thus, entities closer in proximity have higher similarity values. Our defined similarity measure between time series $X_i(t)$ and $X_j(t)$, denoted as $s(X_i(t), X_j(t))$, is therefore expressed as the negative distance:
\begin{equation}
s(X_i(t), X_j(t)) = -d(X_i(t), X_j(t)) = -f(X_i(t), X_j(t)),
\label{eq:d_s}
\end{equation}
where $f(X_i(t), X_j(t)))$ signifies the distance function between $X_i(t)$ and $X_j(t)$. This transformation into a similarity measure enables intuitive interpretation and comparison among different pairs of time series. Our framework employs a common distance measure, such as Manhattan, Euclidean, or Mahalanobis distance \cite{Perlibakas2004}, each offering distinct advantages. Manhattan distance, with its simplicity, is beneficial in fMRI analysis where uncorrelated dimensions frequently appear, especially following component decomposition \cite{Khosla2019}. Euclidean distance calculates the direct path, useful for mapping direct signal interactions in complex brain networks \cite{Xiao2022}. Mahalanobis distance measures multidimensional distances considering covariance, beneficial for fMRI scans where neural interactions are intrinsically multidimensional \cite{Khosla2019}.

To improve the effectiveness of our representation for predicting treatment responses, we propose integrating distance-based similarity measures with traditional correlation-based similarity measures. This integration provides complementary perspectives on brain dynamics, addressing limitations such as overlooking connectivity and scaling differences in correlation-based measures \cite{Xiao2022}. Distance-based measures capture these differences, although they may be sensitive to outliers. By combining both types, we aim to achieve a more comprehensive understanding of the data. In our approach, we represent the FC at a given time point $t$ using two matrices: (1) the Pearson correlation coefficient (PCC) matrix $r(t)$ and (2) the negative pairwise distance matrix $d(t)$ proposed in equation \ref{eq:d_s}. These matrices are computed as follows:

\begin{equation}
r_{ij}(t) = \frac{Cov(X_i(t), X_j(t))}{\sigma_{X_i(t)}\sigma_{X_j(t)}} \in \mathbb{R}^{N\times N},
\end{equation}

\begin{equation}
d_{ij}(t) = -d(X_i(t), X_j(t)),
\end{equation}
where $Cov(X_i(t), X_j(t))$ denote the covariance between the $i$-th and $j$-th ROI at $t$, while $\sigma_{X_i(t)}$ and $\sigma_{X_j(t)}$ denote the corresponding standard deviations.

\noindent \textbf{Dynamic Correlation-Distance Graph Representation} 
We introduce a dynamic graph representation design to encapsulate correlation and distance-based similarity that characterize the dynamic FC within a single brain scan. In contrast to prior work that employed static FC representations using distance-based similarity, we utilize a dynamic FC graph representation for rs-fMRI analysis \cite{Xiao2022,Ma2023}. The construction of this dynamic graph enables us to elucidate varying associations and connections among brain regions and subregions, thereby representing the topological relationships of ROIs within the broader brain network \cite{Smitha2017}. It is particularly effective in decoding brain activities considering the non-stationary dynamic fluctuations inherent to the brain \cite{Chang2010,Thompson2018}.

Specifically, we employ a sliding window approach to extract sub-sequences $X(t)$ from the original signal sequence $X$ \cite{Babcock2002}, where the window setting is adjusted to different scenarios and overlapping is allowed. This approximates evolving FC patterns and preserves neural correlates at shorter scales \cite{Thompson2018}. We assemble a brain FC network graph $G(V, E)$ \cite{Wang2010}, using ROIs as nodes $V$ and FC between ROIs as edges $E$. This graph captures the FC temporal evolution at specific time windows $t$ as $G(t)(V(t), E(t))$, where $V(t)$ and $E(t)$ respectively represent the vertices and edges at a time window $t$. Based on previous experimental study \cite{Kim2020}, binary adjacency matrices $A_{r}(t)  \in \{0, 1\}$ and $A_{d}(t) \in \{0, 1\}$ are derived from the correlation and distance-based neural similarity by thresholding the top $30\%$ percentile values of the FC matrices $r(t)$ and $d(t)$, resulting in sparse graphs that provide a clear FC representation for modeling.

To capture the temporal patterns of brain activity within $X(t)$, we derive hidden features from the ROI signals of $X(t)$ for our graph. These features are then concatenated with the identity matrix $e_{V} \in \{0, 1\}^{M\times M}$, providing a one-hot encoding for each node. This approach introduces a node-aware constraint on feature modeling, thereby enhancing the depiction of temporal patterns in the dynamic FC \cite{Kim2021}. The temporal hidden feature is formalized as:
\begin{equation}
G(t) = W_{M}[e_{V}||LSTM(X(t))],
\end{equation}
where $LSTM(X(t)) \in \mathbb{R}^{D}$ signifies a Long Short-Term Memory (LSTM) \cite{Hochreiter1997} unit that processes the encoded ROI signals up to the endpoint of each $X(t)$ as input. Here, $W_{M}\in \mathbb{R}^{D \times (M+D)}$ denotes a learnable parameter matrix that is utilized for transforming node features.

\subsection{Correlation-Distance Graph Isomorphism Network}

\noindent \textbf{Graph Isomorphism Network Incorporating Contrastive Loss} We present the CD-GIN with contrastive loss for enhanced treatment response prediction. GINs, efficient in capturing global and local graph features, are utilized for learning correlation and distance-based FC graph features \cite{Kim2020}. Our method diversifies GIN expressive power by incorporating contrastive loss, pushing the model to discern unique features across FC graph streams over time windows. Contrasting with prior fMRI contrastive learning designs \cite{Wang2022}, which primarily focus on learning unique features for each patient, our work emphasizes the differentiation of underlying characteristics within a single patient.  It aligns our graph hidden features with the natural structure of these neural and hemodynamic sources, reflecting their typical temporal autocorrelation relationships \cite{Olszowy2019}. This design enhances the interpretability and performance of our contrastive design.

In detail, we assume a perfect graph learning model $f(X(t), A(t))$ to capture characteristic features that can infer treatment response. Here, $A(t)$ denotes the FC between ROIs within $X(t)$, which is either $A_{r}(t)$ or $A_{d}(t)$ for correlation-based or distance-based FC streams, respectively. To simulate the autocorrelation pattern in fMRI \cite{Olszowy2019}, the feature representations of $P(f(X(t)), A(t))$ and $P(f(X(t\pm\delta)), A(t\pm\delta))$ should be more similar for the same FC $A(t)$ within near time windows $t$ and $t\pm \delta$, where $P$ is a Multi-Layer Perception (MLP) projection with shared weights.  Conversely, considering the unique feature representation of temporal patterns, for non-near time windows $t$ and $t+\Delta$, where $\Delta > \delta$, the feature representations of patterns are dissimilar between $P(f(X(t)),A(t)) $ and $P(f(X(t\pm\Delta)), A(t\pm \Delta))$. Considering the unique representation between $A(t)$ and different FC $\overline{A(t)}$,  the feature representations of patterns should be also dissimilar between $P(f(X(t)),A(t)) $ and $P(f(X(t)),\overline{A(t)}) $, $P(f(X(t\pm\delta)),\overline{A(t\pm\delta)}) $ or $P(f(X(t\pm\Delta)),\overline{A(t\pm\Delta)})$. To achieve this ideal $f(X(t), A(t))$ during training, we treat hidden features from nearby time points within the same graph representation as a positive pair, and those from distant time points or different graph representations as a negative pair. The loss function $L_{info}$ that encourages this assumption can be formulated as follows \cite{Chen2020}:
\begin{equation}
L_{\text{info}} = -\frac{1}{N}\sum_{i=1}^{N}\left[\log\frac{\exp(\text{sim}(z_i, z_{i\pm\delta}))}{\sum_{j}^{N}\exp(\text{sim}(z_i, \overline{z_{j}})) + \sum_{j\neq i\pm\delta}^{N}\exp(\text{sim}(z_i, z_j))}\right]
\end{equation}
where $N$ is the number of time windows, $sim(\cdot)$ denotes the cosine similarity,  $z_i = P(f(X(t_i), A(t_i))$ is the projection of the output of our graph learning model for time point $t_i$, $z_j = P(f(X(t\pm \Delta), A(t\pm \Delta))$ is the projection of the output of our graph learning model for far time point $t_j$ and $\overline{z_j} = P(f(X(t_j), \overline{A(t_j)})$ is the projection of the output of our model for other FC graph representation. 

During the execution of our network, we treat each FC graph stream as $A(t)$, while the alternate stream is considered as $\overline{A(t)}$. Both streams are subject to regulation by our proposed contrastive loss. The hidden features $H_{r}(t)$ and $H_{d}(t)$ for the correlation and distance streams of our ideal model are updated according to the GIN process as follows:
\begin{equation}
H_{r}(t) = R_{r}(MLP_{r}((\epsilon_{r} \cdot I + A_{r}(t))H_{r}^{input}(t)W_{r}),
\end{equation}
\begin{equation}
H_{d}(t) = R_{d}(MLP_{d}((\epsilon_{d} \cdot I + A_{d}(t))H_{d}^{input}(t)W_{d}),
\end{equation}
where $R$ represents an attention-based function for computing the graph's overall representation \cite{Kim2021}, $\epsilon$ is an initially zero parameter that can be learned, $I$ is the identity matrix, and $W$ represents the network weights of the MLP. The subscripts $r$ and $d$ denote the correlation and distance streams, respectively. This structure allows us to capture more non-linear characteristics of functional connectivity for each stream \cite{Xu2018}, thereby increasing the capacity of our network.

\noindent \textbf{Convolutional Block Attention Module}  We integrated the CBAM \cite{Woo2018} into our model to enhance its comprehension at each time point by assigning weights to different graph features corresponding to different time windows. Unlike conventional graph integration methods such as later fusion by average pooling \cite{Dwivedi2022}, the CBAM module enables our model to dynamically adjust its attention focus on critical FC graph representations. Compared to conventional attention, CBAM provides more precise control over the importance assignment to correlation and distance-based FC measures, leading to improved model performance \cite{Woo2018}. This adaptive mechanism captures temporal fluctuations in neural activity and selects informative graph representations, improving accuracy in predicting treatment response in rs-fMRI. It also offers insights into the specific time window and FC graph that contribute to the treatment outcome, valuable for further research. The implementation details are described below:
\begin{equation}
Attn_{Stream} = \sigma(W_{f}(MaxPool(H_{f}(t))) + W_{f}(AveragePool(H_{f}(t))))
\end{equation}
\begin{equation}
Attn_{Temporal} = \sigma(Conv1D(MaxPool(H_{f}(t)))+ Conv1D(AveragePool(H_{f}(t)))),
\end{equation}
where ${H}_{f}(t)$ is $[H_{r}(t) ||H_{d}(t)]$, $\sigma$ is an activation function, $W_{f}$ denotes the network weights of the two-layer MLP for max pooling and average pooling output. The attended hidden features ${H}_{a}(t)$ are obtained by:
\begin{equation}
{H}_{a}(t) = {H}_{f}(t) \times Attn_{Stream} \cdot Attn_{Temporal}
\end{equation}

The final probability of treatment response prediction is inferred from a two-layer MLP based on the concatenation of $k$ layers attended hidden features. The training loss $L$ is a binary cross-entropy with the contrastive loss:
\begin{equation}
L = -y^{T}log(\hat{y}) + \alpha L_{info},
\end{equation}
where $\alpha$ denotes a hyper-parameter to tune the contrastive loss.

\section{Experiment and Results}
\begin{table}[h]
\scriptsize
    \centering
        \caption{Hyper-parameters for Models in Our Experiment. BS: Batch Size; LR: Learning Rate; WD: Weight Decay Rate; WS: Window Size; SS: Stride Size. To determine the optimal values for these hyperparameters, we used Weight \& Bias to select best configurations.}
\begin{tabular}{c|c|c|c|c|c|c|c|c|c|c|c|c|c|c|c|c|c|c|c}
\hline
\multicolumn{2}{c|}{}&\multicolumn{6}{c|}{Duloxetine}&\multicolumn{6}{c|}{Placebo}&\multicolumn{6}{c}{DPD Treatment}\\
\hline
\multicolumn{2}{c|}{Method}&Layer&BS& LR& WD & WS & SS &Layer&BS& LR& WD & WS & SS&Layer&BS& LR& WD & WS & SS\\
\hline
\multicolumn{2}{c|}{LSTM \cite{Hochreiter1997}}&2&2&3e-4&1.5e-4&-&-&2&4&3e-4&3e-5&-&-&2&2&4e-4&4e-5&-&-\\
\multicolumn{2}{c|}{GCN\cite{Kipf2016}}&2&2&2e-4&1e-5&-&-&2&2&1e-4&1e-5&-&-&2&2&3e-4&1.5e-4&-&-\\
\multicolumn{2}{c|}{GIN\cite{Xu2018}}&4&2&5e-4&2.5e-4&-&-&3&2&5e-4&5e-5&-&-&3&2&4e-4&2e-4&-&-\\
\multicolumn{2}{c|}{STGCN\cite{Gadgil2020}}&2&2&1e-4&5e-5&50&30&2&2&3e-4&3e-5&30&10&2&2&2e-4&2e-5&40&9\\
\multicolumn{2}{c|}{STAGIN\cite{Kim2021}} &2&2&4e-4&2e-4&25&5&4&2&2e-4&2e-5&50&5&4&2&5e-4&2.5e-4&50&7\\
\hline
\multicolumn{2}{c|}{Ours} &2&4&4e-4&2e-4&35&25&4&2&4e-4&4e-5&50&5&2&2&4e-4&4e-6&35&20\\
\hline
    \end{tabular}
    \label{tab:hyper}
\end{table}
\subsection{Experimental Design}
\noindent \textbf{Dataset}
Our study used a public clinical trial dataset (\url{https://openfmri.org/s3-browser/?prefix=ds000208}, \cite{Tetreault2016}) and a custom dataset, DPD45. The public dataset has pre-treatment fMRI data from 54 OS patients (8 responders and 9 non-responders to duloxetine; 18 responders and 19 non-responders to placebo). The DPD45 dataset, built upon our previous study \cite{Zheng2022}, has pre-treatment fMRI data from 45 DPD patients (19 responders, 26 non-responders, screened by \cite{Sierra2000}) collected before routine treatment. Data preprocessing was conducted with the DPABI toolbox \cite{Yan2016}, and the HCP-MMP1 and Harvard-Oxford atlases \cite{Glasser2016,Jenkinson2012} were used for ROI signal extraction. All procedures were approved by our institutional ethics committee, and all participants provided informed consent.

\noindent \textbf{Benchmark Models}
Our benchmark models primarily include state-of-the-art dynamic learning methods for fMRI analysis and other conventional methods used in previous machine learning-based fMRI analyses \cite{Khosla2019}. Models such as LSTM \cite{Hochreiter1997}, GCN \cite{Kipf2016}, GIN \cite{Xu2018}, and the latest DG methods including STGCN \cite{Gadgil2020}, and STAGIN \cite{Kim2021} were used for comparative analysis.

\noindent \textbf{Training Configuration}
The experiments used the Weight \& Bias tool for hyperparameter optimization and the PyTorch 1.9.0 framework on a GTX 2080 Ti GPU server. We tested our model performance with different distance measures for the negative distance neural similarity: Manhattan distance, Euclidean distance, and Mahalanobis distance. 
The $\alpha$ and $\delta$ parameters were set to 0.1 and 1, respectively, in our framework. Each model was trained using the Adam optimizer, and all other hyperparameters are provided in Table \ref{tab:hyper}.
We stratified the dataset into 80$\%$ training and 20$\%$ testing. Each model was appraised via 4-fold stratified cross-validation, maintaining identical hyperparameters.

\begin{table}[h]
\scriptsize
    \centering
        \caption{Performance with OS Dataset. PCC: Pearson Correlation Coefficient stream; MD: Manhattan Distance stream; ED: Euclidean Distance stream; MaD: Mahalanobis Distance stream}
\begin{tabular}{c|c|c|c|c|c|c|c|c|c}
\hline
\multicolumn{2}{c|}{}&\multicolumn{4}{c|}{Duloxetine}&\multicolumn{4}{c}{Placebo}\\
\hline
\multicolumn{2}{c|}{Method}&AUC& ACC & SE & SP &AUC& ACC & SE & SP\\
\hline
\multicolumn{2}{c|}{LSTM \cite{Hochreiter1997}}&0.62$\pm$0.12&0.56$\pm$0.10&0.75$\pm$0.25&0.37$\pm$0.21&0.67$\pm$0.05&0.59$\pm$0.10&0.37$\pm$0.41&\textbf{0.81$\pm$0.32}\\
\multicolumn{2}{c|}{GCN\cite{Kipf2016}}&0.56$\pm$0.27&0.56$\pm$0.10&0.25$\pm$0.43&0.87$\pm$0.21&0.60$\pm$0.05&0.53$\pm$0.05&0.68$\pm$0.40&0.37$\pm$0.41\\
\multicolumn{2}{c|}{GIN\cite{Xu2018}}&0.68$\pm$0.10&0.68$\pm$0.10&0.75$\pm$0.43&0.62$\pm$0.21&0.64$\pm$0.09&0.65$\pm$0.10&0.50$\pm$0.30&\textbf{0.81$\pm$0.10}\\
\multicolumn{2}{c|}{STGCN\cite{Gadgil2020}}&0.62$\pm$0.27&0.68$\pm$0.20&0.62$\pm$0.41&0.75$\pm$0.43&0.62$\pm$0.22&0.59$\pm$0.13&0.68$\pm$0.27&0.43$\pm$0.27\\
\multicolumn{2}{c|}{STAGIN\cite{Kim2021}} &0.75$\pm$0.17&0.75$\pm$0.17&0.62$\pm$0.41&\textbf{0.87$\pm$0.21}&0.70$\pm$0.17&0.62$\pm$0.08&\textbf{0.75$\pm$0.30}&0.50$\pm$0.39\\
\hline
\multirow{3}*{\rotatebox{90}{Ours}}&PCC+MD &0.75$\pm$0.43&0.62$\pm$0.27&0.75$\pm$0.25&0.50$\pm$0.50&0.65$\pm$0.09&0.50$\pm$0.08&0.56$\pm$0.44&0.43$\pm$0.44\\
~&PCC+ED&\textbf{0.93$\pm$0.10}&\textbf{0.81$\pm$0.10}&\textbf{0.87$\pm$0.21}&0.75$\pm$0.25&\textbf{0.76$\pm$0.17}&\textbf{0.71$\pm$0.13}&0.68$\pm$0.10&0.75$\pm$0.17\\
~&PCC+MaD&0.87$\pm$0.12&0.68$\pm$0.20&0.75$\pm$0.43&0.62$\pm$0.41&0.71$\pm$0.10&0.56$\pm$0.13&0.56$\pm$0.36&0.56$\pm$0.27\\
\hline
    \end{tabular}
    \label{tab:pain}
\end{table}

\begin{table}[h]
\scriptsize
    \centering
        \caption{Performance with DPD Dataset.  PCC: Pearson Correlation Coefficient stream; MD: Manhattan Distance stream; ED: Euclidean Distance stream; MaD: Mahalanobis Distance stream}
\begin{tabular}{c|c|c|c|c|c}
\hline
\multicolumn{2}{c|}{Method} &AUC& ACC & SE & SP\\
\hline
\multicolumn{2}{c|}{LSTM \cite{Hochreiter1997}}&0.66$\pm$0.05&0.58$\pm$0.04&0.68$\pm$0.20&0.50$\pm$0.22\\
\multicolumn{2}{c|}{GCN\cite{Kipf2016}}&0.71$\pm$0.16&0.61$\pm$0.09&0.25$\pm$0.43&\textbf{0.90$\pm$0.17}\\
\multicolumn{2}{c|}{GIN\cite{Xu2018}}&0.58$\pm$0.09&0.61$\pm$0.09&0.37$\pm$0.12&0.80$\pm$0.14\\
\multicolumn{2}{c|}{STGCN\cite{Gadgil2020}}&0.70$\pm$0.18&0.58$\pm$0.09&0.50$\pm$0.39&0.65$\pm$0.21\\
\multicolumn{2}{c|}{STAGIN\cite{Kim2021}} &0.65$\pm$0.09&0.61$\pm$0.12&0.56$\pm$0.36&0.65$\pm$0.25
\\
\hline
\multirow{3}*{\rotatebox{90}{Ours}}&PCC+MD&0.58$\pm$0.08&0.58$\pm$0.09&0.81$\pm$0.20&0.40$\pm$0.24\\
~&PCC+ED&\textbf{0.71$\pm$0.04}&\textbf{0.72$\pm$0.05}&0.81$\pm$0.20&0.65$\pm$0.16\\
~&PCC+MaD&0.68$\pm$0.07&0.63$\pm$0.09&\textbf{0.87$\pm$0.12}&0.45$\pm$0.21\\
\hline
    \end{tabular}
    \label{tab:dpd}
\end{table}

\begin{table}[ht]
\scriptsize
    \centering
        \caption{Ablation Study with OS Dataset. PCC: Pearson Correlation Coefficient stream; ED: Euclidean Distance stream; CL: Contrastive Loss}
\begin{tabular}{c|c|c|c|c|c|c|c|c|c}
\hline
\multicolumn{2}{c|}{}&\multicolumn{4}{c|}{Duloxetine}&\multicolumn{4}{c}{Placebo}\\
\hline
\multicolumn{2}{c|}{Method}&AUC& ACC & SE & SP &AUC& ACC & SE & SP\\
\hline
\multicolumn{2}{c|}{Full Model}&\textbf{0.93$\pm$0.10}&\textbf{0.81$\pm$0.10}&\textbf{0.87$\pm$0.21}&0.75$\pm$0.25&\textbf{0.76$\pm$0.17}&\textbf{0.71$\pm$0.13}&0.68$\pm$0.10&\textbf{0.75$\pm$0.17}\\
\multicolumn{2}{c|}{w/o PCC}&0.50$\pm$0.35&0.50$\pm$0.12&0.37$\pm$0.41&0.37$\pm$0.41&0.62$\pm$0.13&0.59$\pm$0.10&0.68$\pm$0.40&0.50$\pm$0.35\\
\multicolumn{2}{c|}{w/o ED}&0.75$\pm$0.30&0.62$\pm$0.12&0.62$\pm$0.41&0.62$\pm$0.41&0.60$\pm$0.20&0.56$\pm$0.13&0.62$\pm$0.37&0.50$\pm$0.35\\
\multicolumn{2}{c|}{w/o CL}&0.87$\pm$0.21&0.68$\pm$0.10&0.50$\pm$0.35&\textbf{0.87$\pm$0.21}&0.59$\pm$0.03&0.53$\pm$0.05&\textbf{0.81$\pm$0.20}&0.25$\pm$0.25\\
\multicolumn{2}{c|}{w/o CBAM}&\textbf{0.93$\pm$0.10}&0.75$\pm$0.25&0.75$\pm$0.43&0.75$\pm$0.43&0.68$\pm$0.22&0.53$\pm$0.10&0.75$\pm$0.25&0.31$\pm$0.10\\
\hline
    \end{tabular}
    \label{tab:ab_pain}
\end{table}

\begin{table}[h]
\scriptsize
    \centering
        \caption{Ablation Study with DPD Dataset. PCC: Pearson Correlation Coefficient stream; ED: Euclidean Distance stream; CL: Contrastive Loss}
\begin{tabular}{c|c|c|c|c|c}
\hline
\multicolumn{2}{c|}{Method} &AUC& ACC & SE & SP\\
\hline
\multicolumn{2}{c|}{Full Model}&\textbf{0.71$\pm$0.04}&\textbf{0.72$\pm$0.05}&\textbf{0.81$\pm$0.20}&\textbf{0.65$\pm$0.16}\\
\multicolumn{2}{c|}{w/o PCC}&0.46$\pm$0.06&0.50$\pm$0.05&0.56$\pm$0.27&0.45$\pm$0.16\\
\multicolumn{2}{c|}{w/o ED}&0.55$\pm$0.08&0.52$\pm$0.04&0.56$\pm$0.10&0.50$\pm$0.10\\
\multicolumn{2}{c|}{w/o CL}&0.58$\pm$0.04&0.52$\pm$0.04&0.62$\pm$0.21&0.45$\pm$0.16\\
\multicolumn{2}{c|}{w/o CBAM}&0.55$\pm$0.10&0.58$\pm$0.04&0.75$\pm$0.17&0.45$\pm$0.08\\
\hline
    \end{tabular}
    \label{tab:ab_dpd}
\end{table}

\begin{table}[h]
\scriptsize
    \centering
    \caption{Sensitive Analysis of $\alpha$ and $\delta$ on Accuracy}
    \begin{tabular}{c|c|c|c}
         Parameter& Duloxetine& Placebo& DPD Routine Treatment \\
    \hline
    $\alpha = 0.1, \delta = 1$  & \textbf{0.81$\pm$0.10} &\textbf{0.71$\pm$0.13} & \textbf{0.72$\pm$0.05}\\
    $\alpha = 0.1, \delta = 2$  & 0.75$\pm$0.17&0.40$\pm$0.10 &0.58$\pm$0.04 \\
    $\alpha = 0.1, \delta = 3$  & 0.65$\pm$0.12&0.53$\pm$0.13 &0.55$\pm$0.07 \\
    $\alpha = 0.5, \delta = 1$  &\textbf{0.81$\pm$0.10}& 0.50$\pm$0.08 & 0.55$\pm$0.07\\
    $\alpha = 0.5, \delta = 2$  &0.75$\pm$0.17&0.59$\pm$0.16 & 0.61$\pm$0.09\\
    $\alpha = 0.5, \delta = 3$  & \textbf{0.81$\pm$0.10} & 0.62$\pm$0.15&0.55$\pm$0.05 \\
    $\alpha = 1.0, \delta = 1$  & 0.65$\pm$0.27& 0.53$\pm$0.10& 0.58$\pm$0.04\\
    $\alpha = 1.0, \delta = 2$  &0.68$\pm$0.10 &0.50$\pm$0.00 &0.52$\pm$0.04\\
    $\alpha = 1.0, \delta = 3$  &\textbf{0.81$\pm$0.20} &0.50$\pm$0.00  &0.58$\pm$0.04 \\
    \hline
    \end{tabular}
    \label{tab:sen}
\end{table}

\begin{figure*}[h]
\centering
\includegraphics[scale = 0.22]{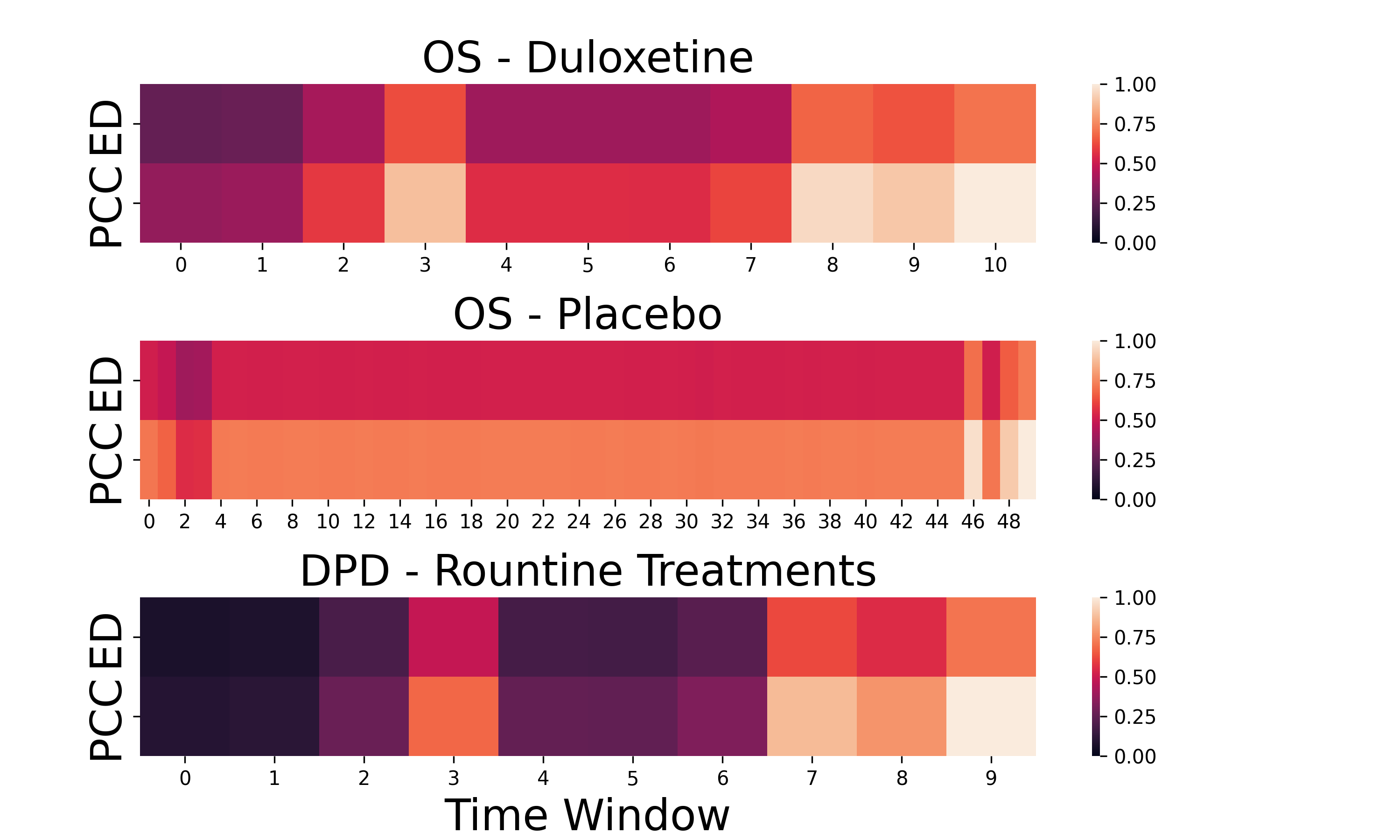}
\caption{Attention Map ($Attn_{Stream} \cdot Attn_{Temporal}$) of PCC vs. ED by Time Windows}
\label{fig:Att}
\end{figure*}

\noindent \textbf{Metrics}
Performance was evaluated by various metrics, including the area under the ROC curve (AUC), accuracy (ACC), sensitivity (SE), and specificity (SP), according to previous studies for treatment response prediction \cite{Kesler2017}.

\subsection{Results and Discussion}
\noindent \textbf{Quantitative Analysis}
Our model exceeds baseline methods in predicting treatment outcomes as shown in Table \ref{tab:pain} and Table \ref{tab:dpd}. Though sensitivity or specificity may appear lower in some scenarios, it is crucial to note the data imbalance and treatment complexities. For instance, placebo administration and DPD routine treatments vary in duration and are individually tailored, unlike the consistent duloxetine treatment. Nonetheless, our findings highlight the wide applicability of our model in diverse scenarios, including clinical trials and real-world clinical practice. Throughout our experimentation with different distance measures, we consistently observe that the Euclidean distance-based similarity achieves superior results. This can be attributed to its direct assessment of overall dissimilarity between ROIs, without relying on specific assumptions. This finding suggests that Euclidean distance is the most suitable metric for constructing distance-based neural similarity within our framework.

In our ablation study, we used the Euclidean distance as the basis for distance-based neural similarity due to its superior performance. The results, shown in Tables \ref{tab:ab_pain} and \ref{tab:ab_dpd}, confirm that our model, including all components, achieves balanced performance in both OS and DPD datasets. Notably, the AUC and ACC metrics highlight the model effectiveness. That supports our hypothesis regarding the complementary relationship between correlation and distance-based streams in predicting treatment responses. Importantly, excluding either the correlation or distance stream resulted in varying degrees of performance decline, suggesting the influence of dataset complexity. This demonstrates the adaptability of our approach across different data domains and provides evidence that incorporating both correlation and distance-based streams enhances performance.

Our sensitivity analysis (Table \ref{tab:sen}) investigates the impact of parameters $\alpha$ and $\delta$ on model accuracy. The model maintains strong performance across different $\alpha$ and $\delta$ values for duloxetine, indicating robustness. However, in placebo and DPD routine treatments, optimal accuracy is achieved at $\alpha = 0.1, \delta = 1$, with a decrease as $\delta$ rises, demonstrating the model sensitivity to $\delta$ in these cases. This shows the model adaptability to situations with consistent interventions, although in more complex cases, hyperparameter fine-tuning may be required.

\noindent \textbf{Qualitative Analysis}
Figure \ref{fig:Att} shows the attention values of graphs derived from PCC and ED streams across various time windows in our test set. The attention values indicate the importance of each stream at different time points. Despite PCC stream having generally higher values, the ED stream also plays a significant role. This validates our ablation study results, suggesting complementary roles of PCC and ED in treatment prediction. Moreover, differences in time windows and attention distribution between duloxetine and placebo show the varied brain dynamics between treatments, highlighting the interpretability and dynamic nature of our method.

\section{Conclusion}
In this work, we introduce a graph learning framework combining correlation-based and distance-based neural similarity to enhance treatment response prediction understanding. Our validation on two datasets highlights Euclidean distance as particularly effective for distance-based FC, showing the potential of our model for real-world clinical application. Future research will refine our theoretical basis through clinical trials, examining the impact of atlas and FC threshold choices. Additionally, we will also explore advanced methodologies like diffusion models to infer deeper brain mechanisms for treatment response \cite{Chang2023}, and investigate the applicability of our framework to other clinical contexts requiring complex graph representation, such as surgery \cite{Zhang2022}.

%
%
%
%

\end{document}